\documentclass[reprint,aps,prb,superscriptaddress,showpacs]{revtex4-1}
\usepackage{graphicx}
\usepackage{amsmath}
\usepackage{amsfonts}
\usepackage{amssymb}
\usepackage{srcltx}
\usepackage{CJK}
\usepackage[breaklinks=true,colorlinks=true,linkcolor=blue,citecolor=blue,urlcolor=black,bookmarks=false]{hyperref}

\DeclareMathOperator{\Var}{Var}

\begin{document}

\begin{CJK*}{Bg5}{bsmi}
\title{Spin Conductance of Diffusive Graphene Nanoribbons:\\ A Probe of Zigzag Edge Magnetization}%

\author{Jan Bundesmann}
\affiliation{Institut f\"{u}r Theoretische Physik, Universit\"{a}t Regensburg, 93040 Regensburg, Germany}%
\author{Ming-Hao Liu (¼B©ú»¨)}
\affiliation{Institut f\"{u}r Theoretische Physik, Universit\"{a}t Regensburg, 93040 Regensburg, Germany}%
\author{\.{I}nan\c{c} Adagideli}
\affiliation{Faculty of Engineering and Natural Sciences, Sabanci University, 34956 Orhanli-Tuzla, Istanbul, Turkey}
\author{Klaus Richter}
\affiliation{Institut f\"{u}r Theoretische Physik, Universit\"{a}t Regensburg, 93040 Regensburg, Germany}%
\pacs{73.23.-b 72.25.-b 72.80.Vp}
\begin{abstract}
We investigate spin transport in diffusive graphene nanoribbons with both clean and rough zigzag edges, and long-range potential fluctuations. The long-range fields along the ribbon edges cause the local doping to come close to the charge neutrality point forming $p$-$n$ junctions with localized magnetic moments, similar to the predicted magnetic edge of clean zigzag graphene nanoribbons. The resulting random edge magnetization polarizes charge currents and causes sample-to-sample fluctuations of the spin currents obeying universal predictions. We show furthermore that, although the average spin conductance vanishes, an applied transverse in-plane electric field can generate a finite spin conductance. A similar effect can also be achieved by aligning the edge magnetic moments through an external magnetic field.
\end{abstract}

\maketitle

\end{CJK*}

\section{Introduction}
Understanding graphene edges is of profound interest for the investigation of graphene nanostructures as the edges induce peculiar features depending on the edge orientation, that strongly influence the electronic properties of graphene. 
Armchair nanoribbons can be either metallic or semiconducting depending on the ribbon width while zigzag ($zz$) nanoribbons are always metallic due to the presence of a state localized at the edge \cite{JPSJ-65-1920,PhysRevB.54.17954}.
Moreover the $zz$ edges are predicted to be magnetic at half filling, with oppositely spin-polarized edges, based on the mean-field approximation of the Hubbard \cite{JPSJ-65-1920} and the extended Hubbard model \cite{PhysRevB.68.193410}.
Density functional theory (DFT) calculations \cite{son_half-metallic_2006,PhysRevLett.100.177207,PhysRevLett.102.157201,PhysRevLett.106.236805}, exact diagonalization and quantum Monte Carlo simulations \cite{PhysRevB.81.115416}, as well as diagrammatic perturbation theory \cite{PhysRevB.86.115446}, produced results affirming the $zz$ edge magnetization.
On the other hand the stability of the edge state (and consequently its magnetization) has been doubted, as it should only exist for edges passivated with a single hydrogen atom and, moreover, the magnetic ordering should only be observable under ``ultraclean, low-temperature conditions in defect-free samples'' \cite{PhysRevB.83.045414}.
For a recent review see Ref. \onlinecite{0034-4885_73_5_056501}.

On the experimental side, scanning tunneling microscopy (STM) measurements indeed confirmed the presence of an increased local density of states at $zz$ edges \cite{PhysRevB.73.125415,Ritter2009,nn303730v}.
The measurements, however, were not spin sensitive.
While there is evidence of magnetism in graphene \cite{jp903397u,PhysRevB.81.245428,nphys2183}, its origin might also be adatoms and impurities in addition to magnetic $zz$ edges.

\begin{figure}[b]
\centering
\includegraphics{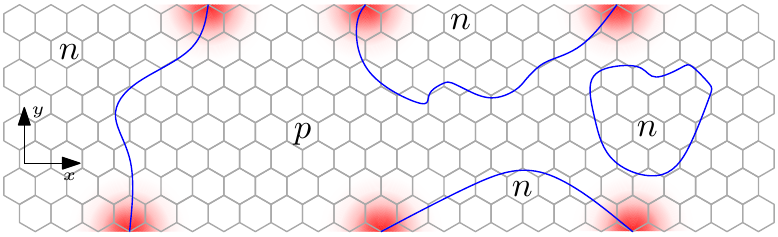}
\caption{Formation of local zigzag edge magnetic moments. Whenever isopotential lines (blue) separating $n$-doped from $p$-doped regions hit an edge, a finite magnetization is locally assumed (red).}
\label{fig:sketch_magnetization}
\end{figure}

Theoretical studies already suggested measuring spin transport as a probe of 
edge magnetism \cite{raey,Farghadan2012}. However, it was unclear whether the predictions, which 
assumed ribbons close to half-filling, were directly applicable to experiments 
where both edge roughness and disorder could potentially mask the effect of edge 
magnetism. In this work, we show that disordered graphene nanoribbons exhibit 
universal spin conductance fluctuations due to edge-magnetism. Moreover, the 
spin conductance fluctuations remain finite in a large energy interval as long 
as potential disorder induces charge neutrality at the edges. Furthermore, we 
propose how to control the spin conductance electrically.

\section{Model}
In realistic systems, long-range potential fluctuations generate lines of local charge neutrality. We assume that magnetic clusters form where these lines and zz edges coincide; see Fig.\ \ref{fig:sketch_magnetization}. 
In order to check this assumption, we self-consistently calculated the mean-field Hubbard Hamiltonian for simple systems with non-constant potential and found that local magnetic moments indeed formed at zz edges near charge-neutrality points.
Further sources of magnetic moments in graphene were ignored. Defects or non-magnetic adatoms could, possibly, also induce magnetism in graphene close to charge-neutrality. The probability of a point-like defect coinciding with local charge neutrality in a disordered systems is, however, small compared to the probability of a sequence of edge atoms with a local potentential close to charge-neutrality. We considered systems free of magnetic impurities as their deposition, nowadays, is well controlled \cite{PhysRevLett.109.186604}.
We now consider spin-dependent quantum transport through disordered $zz$ nanoribbons and analyze imprints of local $zz$ edge magnetism in the spin conductance for various scenarios of the relative orientation of the magnetic clusters.
We use a tight-binding description of graphene,
\begin{align}
\begin{split}
\mathcal{H} =& \sum_{i,s} V(\vec{r}_i) c_{i,s}^{\dagger} c_{i,s} + \sum_{i,j,s} t_{i,j} c_{i,s}^{\dagger} c_{j,s} \\
&+ \sum_{i,s, s'} c_{i,s}^{\dagger} (\vec{m}_i \cdot \vec{\sigma})_{s,s'} c_{i,s'}, 
\end{split}\label{eqn::hamiltonian}
\end{align}
where the sums run over atomic sites ($i$ and $j$) and spin indices ($s$ and $s'$). Here $t_{i,j} = t$ if the atoms $i$ and $j$ are nearest neighbors and $t_{i,j} = t'$ for next-nearest neighbors. 
$\vec{\sigma} = (\sigma_x, \sigma_y, \sigma_z)$ is the vector of Pauli matrices in real spin space, thus the third term in Eq. \eqref{eqn::hamiltonian} acts like a Zeeman term. 


We now specify our model. To model long range disorder that may originate from, say, charges trapped in the substrate, we shall adopt a smooth static disorder potential, given by
\begin{align}
V(\vec{r}) &= \sum_n V_n \exp\left(- \frac{\lvert \vec{r} - \vec{r}_n \rvert^2}{2 \sigma_{\text{dis}}^2}\right),\label{eqn:disorder}
\end{align}
that is characterized by a random disorder strength $V_n \in [-V_{\mathrm{dis}}, V_{\mathrm{dis}}]$ and a correlation length $\sigma_{\mathrm{dis}}$. 
\begin{figure}[b]
\centering
\includegraphics{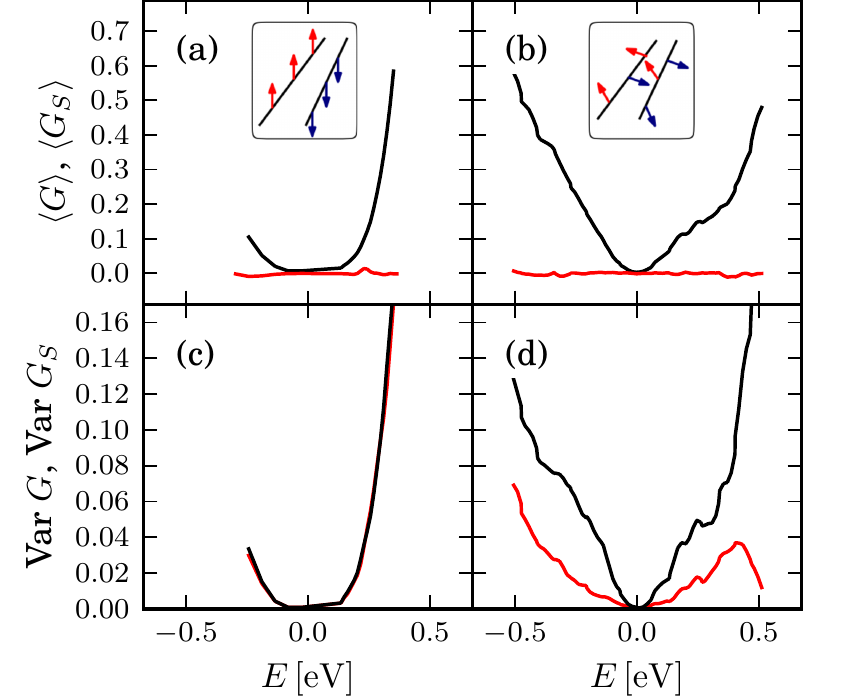}
\caption{Energy dependence of average conductance $\langle G \rangle$ (black) in units of ${e^2}/{h}$ and spin conductance $\langle G_S \rangle$ (red) in units of $\frac{e}{4 \pi}$ for (a) model \textit{a} and (b) model \textit{b} (see text). Corresponding variance of charge and spin conductance for model \textit{a} (c) and \textit{b} (d).}\label{fig::malu_energy_dependence}
\end{figure}

Owing to the potential fluctuations we assume the magnetization to be finite only near $p$-$n$ junctions close to the edge 
\footnote{A $p$-$n$ junction, in this context, is given when the local potential fulfills $- t' \leq E - V(\vec{r}) \leq 0$. These are the points where an increased local density of states exists similar to the $zz$ edge state, which then according to the Hubbard model leads to an ordered phase. We also assume the atomic chain to have a minimum length of 3 atoms that an edge state is formed according to Ref. \onlinecite{PhysRevB.54.17954}.}.
The magnetization is further assumed to decay with distance $d$ from the $p$-$n$ junction at the edge as a Gaussian, $\exp(- d^2 / d_0^2)$, where $d_0$ is a phenomenological decay length.
Several magnetic clusters will form along the edges; see Fig.\ \ref{fig:sketch_magnetization}.
Within these clusters the magnetization has opposite sign for sublattice $A$ and $B$, but the net magnetization is nevertheless finite.

\begin{figure}[b]
\centering
\includegraphics{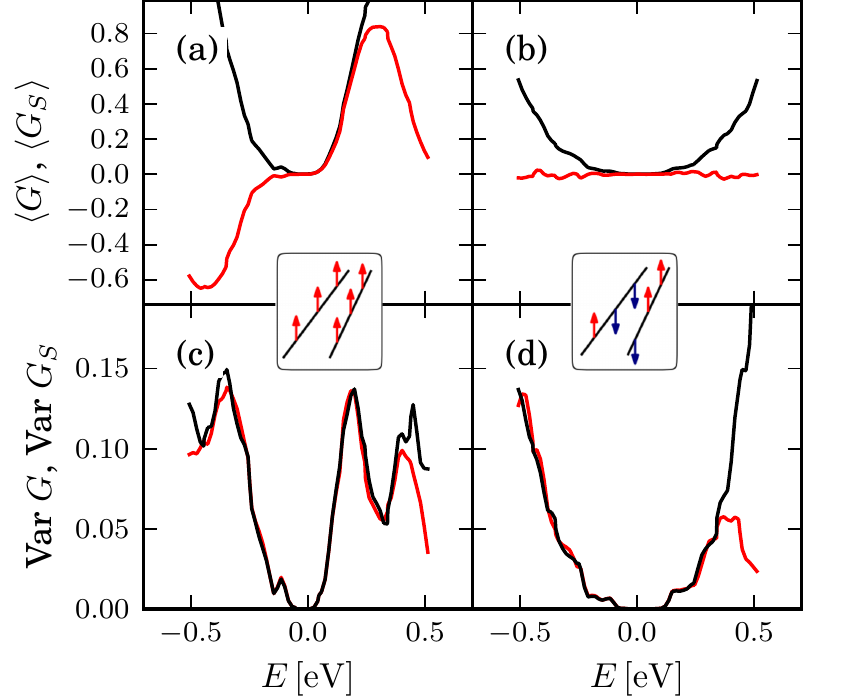}
\caption{$\langle G \rangle$ (black) in units of ${e^2}/{h}$ and $\langle G_S \rangle$ (red) in units of ${e}/{4 \pi}$ as a function of energy $E$ for (a) ferromagnetically aligned magnetic clusters in a rather clean ribbon according to model \textit{c} and (b) when edge disorder randomizes the $A$ and $B$ sublattice segments and thus the sign of the magnetic moments. (c, d) Corresponding variance of $G$ and $G_S$.}\label{fig::extB_energy_dependence}
\end{figure}

We consider three models in this paper how the different clusters relatively align; see also the insets in Figs.\ \ref{fig::malu_energy_dependence} and \ref{fig::extB_energy_dependence}.
\paragraph{Fully Aligned Moments.}
This configuration is an extension of the simple model of constant magnetization along a ribbon's edge which has been used before to model transport in graphene \cite{son_half-metallic_2006,PhysRevLett.100.177207}.
If the magnetic cluster formed at an edge segment is mainly composed of atoms of sublattice $A$ ($B$) its net magnetization points upwards (downwards).
\paragraph{Uncorrelated Moments.}
The mean-field description used to derive the antiferromagnetic alignment of the sublattices of graphene is not able to determine a preferred axis along which the electrons' spins align. 
Magnetic moments at different $p$-$n$ junctions need not be aligned nor collinear. 
Thus it is a natural and realistic extension of model \textit{a} to assign a random direction to the magnetization at each $p$-$n$ junction.
\paragraph{Ferromagnetic ordering of the edges.}
This model is similar to model \textit{a}. The clusters, however, get assigned a direction that additionally depends on the edge they are lying on.
Clusters with mainly edge atoms of sublattice $A$ ($B$) point upwards (downwards) on the left edge and downwards (upwards) on the right edge.
Formally the Hubbard model allows for this solution as shown for clean $zz$ nanoribbons (cf. Fig.\ 4 in Ref. \onlinecite{PhysRevB.79.235433}).
This phase might be triggered by applying small magnetic field.

\section{Transport calculations}
We model transport through graphene nanoribbons in the presence of random long-range disorder [Eq.\ \eqref{eqn:disorder}] with $V_{\text{dis}} = 300 \mathrm{meV}$, $\sigma_{\text{dis}} \approx 1 \mathrm{nm}$.
For model \textit{b}, also the orientation of the magnetic moments at different $p$-$n$ junctions is randomized.
Both, nanoribbons with smooth and rough edges were considered.
Edge disorder is created by iterating over the edge atoms and removing them with a probability of $2 \%$ for this work.
This procedure was repeated up to ten times to increase edge disorder and to extend the size of edge defects.
An example is shown in the inset of Fig.\        \ref{fig::trmalu}.
The edge disorder acts as an additional source of momentum scattering.
It also randomizes the magnetic moments along an edge.
Hence, model \textit{a} and \textit{c} become very similar in the presence of edge disorder; see inset between Figs.\ \ref{fig::extB_energy_dependence}(b) and (d).
Spin-dependent quantum transport is simulated by means of a recursive Green's function method \cite{Wimmer20098548}.
The considered ribbons are $40$ - $50 \mathrm{nm}$ wide and $500 \mathrm{nm}$ - $1 \mathrm{\mu m}$ long.
With these parameters transport takes place mainly in the localized regime.
We calculate spin-resolved transmission probabilities, the transmission matrix in real spin space becomes
\begin{align}
T &= \begin{pmatrix} T_{\uparrow \uparrow} & T_{\uparrow \downarrow} \\ T_{\downarrow \uparrow} & T_{\downarrow \downarrow} \end{pmatrix}.\label{eqn:transmission_matrix}
\end{align}
The charge conductance is the sum of all four probabilities, $G = \frac{e^2}{h} \left( T_{\uparrow \uparrow} + T_{\uparrow \downarrow} + T_{\downarrow \uparrow} + T_{\downarrow \downarrow} \right)$, subtracting transmission to spin down from transmission to spin up, $G_s = \frac{e}{4 \pi} \left( T_{\uparrow \uparrow} + T_{\uparrow \downarrow} - T_{\downarrow \uparrow} - T_{\downarrow \downarrow}\right)$, yields the spin conductance.

\subsection{Average spin conductance}
Without external fields (models \textit{a} and \textit{b}) total average charge conductance $\langle G \rangle$ increases with distance from the CNP, while the average spin conductance $\langle G_S \rangle$ is suppressed; see Figs.\ \ref{fig::malu_energy_dependence}(a, b). 
In the model \textit{a} of fully aligned moments, we find that the sample-to-sample fluctuations of $G$ and $G_S$, i.e. $\Var G$ and $\Var G_S$, coincide; see Fig.\ \ref{fig::malu_energy_dependence}(c).
For model \textit{b} (uncorrelated random directions of magnetic moments) $\langle G_S \rangle = 0$ as expected, but notably the variance is finite; see Fig.\ \ref{fig::malu_energy_dependence}(d). 
$\Var(G_S)$ differs, however, from $\Var (G)$ due to off-diagonal couplings in the transmission matrix \eqref{eqn:transmission_matrix} in this case.

As shown in Fig.\ \ref{fig::extB_energy_dependence}(a), carrying out the transport simulation with ferromagnetically ordered edges (model \textit{c}) yields finite average spin conductance which is positive for positive Fermi energy and vice versa. The extrema of $G_S$ lie at a distance away from the CNP that is consistent with the positions of the spin-split edge states which are not spin-degenerate any more. 

In order to understand the results presented in Figs.\ \ref{fig::malu_energy_dependence} and \ref{fig::extB_energy_dependence}, we consider a simple model in which we assume the magnetic clusters to act as spin filters with a preferred direction depending on the orientation of the localized magnetic moment. 
For a given ribbon we then calculate the number of $p$-$n$ junctions with positive ($N_{\uparrow}$) or negative ($N_{\downarrow}$) orientation, as well as their difference $\Delta = N_{\uparrow} - N_{\downarrow}$, which is proportional to the ribbon's total magnetization 
\footnote{In fact we would have to consider the full three-dimensional magnetization for model \textit{b}. For the sake of simplicity we restrict ourselves to the distinction between spin-up and -down.}.
For models \textit{a} and \textit{b}, $\langle \Delta \rangle = 0$, and hence the average magnetization is zero, resulting in a suppressed average spin transmission.
In model \textit{c}, however, $\Delta$ is finite leading to finite average spin conductance. 
The spin-polarized states are shifted away from the CNP by a value corresponding to the peak strength of the local magnetization \cite{PhysRevB.79.235433}.
This is where spin polarization is most efficient leading to the extrema of the spin conductance.

Within all models, edge disorder can lead to a randomization of the magnetic moments along the edges. 
The effect is negligible for model \textit{a} and \textit{b}, where $\langle \Delta \rangle = 0$. 
In model \textit{c} it leads to decreasing and eventually vanishing $\Delta$.
We find that the average spin conductance also vanishes in agreement with our model; see Fig.\ \ref{fig::extB_energy_dependence}(b).

\subsection{Universal spin conductance fluctuations}
\begin{figure}[t]
\centering
\includegraphics{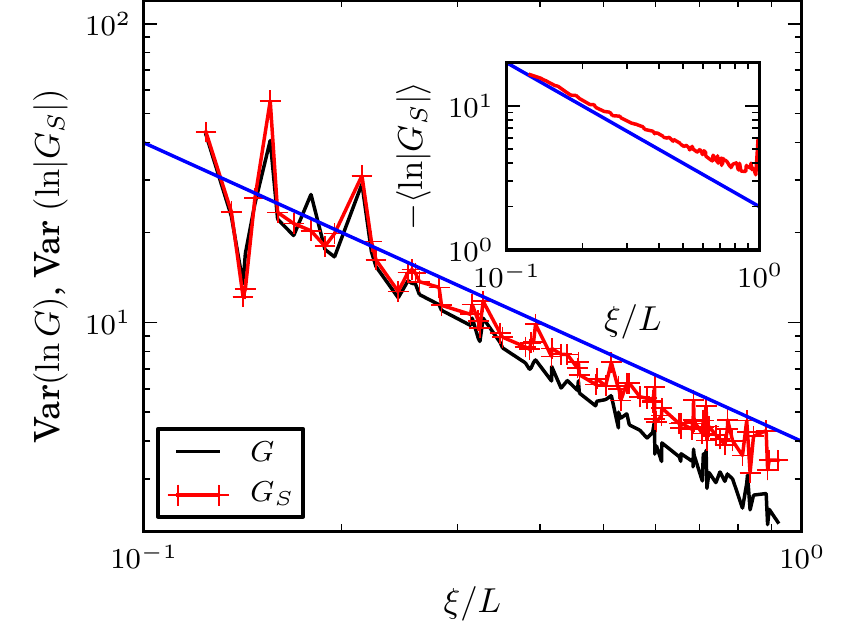}
\caption{$\Var [\ln G]$ for $G$ in units of $\frac{e^2}{h}$ (black) and $\Var [\ln G_S]$ for $G_S$ in units of $\frac{e}{4\pi}$ (red) as a function of $\xi / L$ for various ribbons of length L and localization length $\xi$ and random magnetic moment orientation, model \textit{b}. In the inset the logarithm of the absolute value of the spin conductance is plotted as a function of $\xi / L$ for the same ribbons. The blue lines are the values expected from DMPK equation \cite{Pichard1991}, see text. }\label{fig::comparison_to_rmt}
\end{figure}
An instrument to retrieve general information about mesoscopic systems are conductance fluctuations which, according to random matrix theory, are independent of the particular considered system \cite{RevModPhys.69.731}.
In graphene, they can, e.g., indicate the symmetry class and system degeneracies \cite{PhysRevLett.97.146805,PhysRevB.78.033404,PhysRevLett.102.056806} or be used to extract the phase coherence time \cite{PhysRevB.86.155403}.
Here we focus on sample-to-sample fluctuations.
For strongly localized mesoscopic systems the conductance shows a log-normal distribution \cite{RevModPhys.69.731}.
The calculation of the exact distribution of $\ln G$ for a disordered system is obtained by solving the Dorokhov-Mello-Pereyra-Kumar (DMPK) equations \cite{Dorokhov1982,Mello1988290}.
For the dimensionless conductance, $G / G_0$, with $G_0 = e^2 / h$, this yields $- \langle \ln G / G_0 \rangle = \frac{1}{2} \Var (\ln G / G_0) = 2 L / \xi$ for systems of length $L$ and localization length $\xi$ \cite{Pichard1991}. This prediction is approximately fulfilled for the charge conductance in our simulated ribbons; see black curve in Fig.\ \ref{fig::comparison_to_rmt}.

The spin conductance has to be studied via its absolute value. 
It turns out that, for different systems, mean and variance of $\ln \lvert G_S / (\frac{e}{4 \pi})\rvert$ follow a universal curve as a function of $\xi / L$ independent of the exact choice of the phenomenological parameters describing the local magnetization.
For the aligned magnetic moments (model \textit{a}) $\Var (\ln \lvert G_S / (\frac{e}{4 \pi}) \rvert)$ obeys the same relation as $\Var [\ln (G / G_0)]$ which is an indication that spin-up and -down channel are uncorrelated. 
In Fig.\ \ref{fig::comparison_to_rmt} it can be seen that also for model \textit{b}, uncorrelated magnetic moments, $\Var (\ln \lvert G_S / (\frac{e}{4 \pi}) \rvert)$ follows the same universal law as $\Var [\ln (G / G_0)]$. 
For model \textit{b}, $-\langle \ln \lvert G_S / (\frac{e}{4 \pi}) \rvert \rangle$ is larger than the universal value from DMPK equation; see inset of Fig.\ \ref{fig::comparison_to_rmt}. This is a result of the projection of the three dimensional spin expectation value onto the $z$ axis.
For model \textit{a} such a deviation is not found. 

\subsection{Effect of a transverse in-plane electric field}
For pristine graphene it has been predicted that the application of a transverse in-plane electric field greater than a certain threshold turns a graphene $zz$ ribbon into an insulator for one spin direction and into a metallic phase for the other one \cite{son_half-metallic_2006}.
We now show how an electric field can lead to finite spin conductance in disordered graphene nanoribbons without a threshold for the electric field.
Therefor, we investigate model \textit{a} again in the presence of a potential $V_{\text{tilt}}$ increasing linearly across the ribbon from $-V_0 / 2$ on the left edge to $+V_0 / 2$ on the right edge, which can be viewed as arising from the application of a transverse in-plane electric field.

The obvious effect of a transverse potential drop is a change in the number of $p$-$n$ junctions.
For model \textit{a} without edge disorder this leads to an imbalance between the number of $p$-$n$ junctions on opposite edges and, thereby, directly to the difference $\Delta = N_{\uparrow} - N_{\downarrow}$ between localized magnetic moments pointing upwards and downwards.
As mentioned above, $\Delta \neq 0$ leads to a finite spin conductance. 
\begin{figure}[t]
\centering
\includegraphics{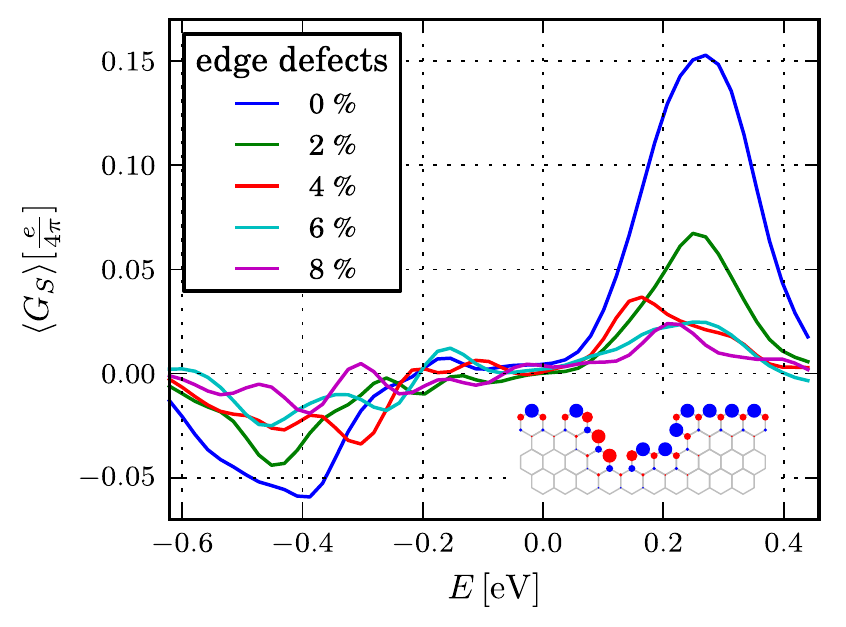}
\caption{Average spin conductance of nanoribbons under the influence of a transverse in-plane electric field. The potential difference leads to a maximum spin conductance at Fermi energies $V_{\text{dis}} - t'$ and $-V_{\text{dis}}$. Different curves show $\langle G_S \rangle$ for increasing edge disorder, from perfectly clean $zz$ edges, to edges with approximately $8 \%$ defects. The inset visualizes how spins align along a defective edge piece with blue (red) circles indicating positive (negative) magnetized atoms. The size of the circles is proportional to the local magnetization.}\label{fig::trmalu}
\end{figure}
A corresponding example is shown in Fig.\ \ref{fig::trmalu}. 
For these calculations magnetization does not exceed $0.01 t$ and decays rather fast, $d_0 = 0.25 \mathrm{nm}$.

In model \textit{a} the transmission as a function of ribbon length $L$, Eq.\ \eqref{eqn:transmission_matrix}, is fully defined by two scaling parameters $L / \xi_{\uparrow}$ and $L / \xi_{\downarrow}$ for the two spin blocks: $T_{\uparrow\uparrow / \downarrow\downarrow} = \exp(-L / \xi_{\uparrow / \downarrow})$, leading to transmission $T = T_{\uparrow\uparrow} + T_{\downarrow\downarrow}$ and spin transmission $T_S = T_{\uparrow\uparrow} - T_{\downarrow\downarrow}$, respectively \footnote{Spin flip is suppressed in model \textit{a}}.
The scaling parameter for each spin block, $L / \xi_{\uparrow / \downarrow}$, depends both on energy and $\Delta$.
To leading order the inverse normalized localization length for each spin block is assumed to depend linearly on $\Delta$,  ${L}/{\xi_{\uparrow / \downarrow}} = {L}/{\xi_0} \pm \gamma \Delta$, as confirmed by our numerical data. This implies $T_S \approx 2 \gamma \Delta \exp(-L / \xi_0)$.
Hence, the positions of the extrema of $T_S(E)$ are given by the peaks of $\Delta(E)$.
For clean zigzag ribbons $\Delta$ is given by the difference of magnetic clusters along left and right edge. 
We assume the local Fermi level for a given transverse coordinate $y$ to exhibit a Gaussian distribution around $E_F - V_{\mathrm{tilt}}(y)$ given by the global Fermi level $E_F$ and the value of the transverse potential drop $V_{\mathrm{tilt}}$ at position $y$. 
The distribution width $\sigma_{E}$ is given by the strength of the long range potential disorder $\sigma_{E} \propto V_{\mathrm{dis}}$. 
Then, the number of $p$-$n$ junctions can be estimated from the energy distribution along the two edges, $\rho_{\text{left/right}}$, and consequently $\Delta(E)$,
\begin{align}
\Delta(E) \propto \int_{-t'}^{0} (\rho_{\text{right}}(E) - \rho_{\text{left}}(E)) \mathrm{d} E.
\end{align}
$\Delta(E)$ is peaked around $\pm \sigma_{E}$ as long as $V_{\mathrm{tilt}} < \sigma_E$ and around $\pm V_{\mathrm{tilt}}$ otherwise. The numerical results shown in Fig.\ \ref{fig::trmalu} follow this prediction.
Notably, if we sharpen the distribution of the local Fermi level by decreasing the disorder strength we eventually recover, as a limiting case, the mechanism of half-metallicity presented in Ref. \onlinecite{son_half-metallic_2006}. 

Edge disorder is expected to reduce $\Delta(E)$ by randomizing the spin orientation along the edges. 
To investigate this effect we simulated transport through nanoribbons at different edge defect rates; see Fig.\ \ref{fig::trmalu}. 
Apparently the spin conductance decreases with increasing edge disorder and tends to zero for a defect rate of about $10 \%$.

While the above proposal could open a route to an all-electric spin current creation and control in graphene, as spin transport experiments are nowadays performed with high accuracy\cite{tombros_electronic_2007,Han2012369}, 
it represents also a generic way of detecting edge magnetization.

\section{Summary}
In conclusion, we considered spin-dependent electron transport in graphene nanoribbons. We showed that even in the diffusive case and for random edge magnetic moments finite spin conductance fluctuations persist following universal predictions. Finite spin conductance fluctuations are visible within a large energy range, demonstrating how potential fluctuations help observing edge magnetism in graphene. Aligning the localized magnetic moments can lead to finite average spin conductance.
Furthermore we showed that the application of a transverse in-plane electrical field can be used to detect edge magnetism and to polarize spin transport in graphene.

\section{Acknowledgements}
We gratefully acknowledge Deutsche Forschungsgemeinschaft (within SFB 689) (J.B. and K.R.), Alexander von Humboldt Foundation (M.H.L.) and the funds of the Erdal \.{I}n\"{o}n\"{u} chair (\.{I}.A.) for financial support. \.{I}.A. thanks University of Regensburg for their hospitality. 

%

\end{document}